\documentclass[aps,pra,twocolumn,superscriptaddress]{revtex4-2}
\usepackage{mathtools}
\usepackage{amsmath,amscd,amssymb,amsthm}

\begin{document}

\title{Tunable linear-optical phase amplification}
\author{Christopher R. Schwarze}
\email[e-mail: ]{{\tt crs2@bu.edu}}
\affiliation{Department of Electrical and Computer Engineering \& Photonics Center, Boston University, 8 Saint Mary’s St., Boston, Massachusetts 02215, USA}
\author{David S. Simon}
\email[e-mail: ]{{\tt simond@bu.edu}}
\affiliation{Department of Electrical and Computer Engineering \& Photonics Center, Boston University, 8 Saint Mary’s St., Boston, Massachusetts 02215, USA}
\affiliation{Department of Physics and Astronomy, Stonehill College, 320 Washington Street, Easton, Massachusetts 02357, USA}
\author{Abdoulaye Ndao}
\email[e-mail: ]{{\tt a1ndao@ucsd.edu}}
\affiliation{Department of Electrical and Computer Engineering \& Photonics Center, Boston University, 8 Saint Mary’s St., Boston, Massachusetts 02215, USA}
\affiliation{Department of Electrical and Computer Engineering,
University of California San Diego, La Jolla, CA, USA}
\author{Alexander V. Sergienko}
\email[e-mail: ]{{\tt alexserg@bu.edu}}
\affiliation{Department of Electrical and Computer Engineering \& Photonics Center, Boston University, 8 Saint Mary’s St., Boston, Massachusetts 02215, USA}
\affiliation{Department of Physics, Boston University, 590 Commonwealth Avenue, Boston, Massachusetts 02215, USA}

\date{\today}
\begin{abstract}
We combine lossless, phase-only transformations with fully-transmitting linear-optical scatterers to define the principle of linear-optical phase amplification. This enables a physical phase shift $\phi$ to be nonlinearly mapped to a new space $\gamma(\phi)$ using linear optics, resulting in a completely general and enhanced phase shifter that can replace any standard one. A particular phase amplifier is experimentally realized, allowing the phase enhancement parameter $d\gamma/d\phi$ to be continuously tuned. Placing this enhanced phase shifter in one arm of a Mach-Zehnder interferometer led to an intensity-phase slope more than twenty times steeper than what can be obtained with its unamplified counterpart.
\end{abstract}

\maketitle

\section{Introduction}
Phase inference and control underlie many applications of optics. Phase information can only be gathered and understood from intensity modulations since these contain the energy that can be measured. The details about these modulations can be used to obtain information about the phase itself. For example, from a small phase perturbation $\Delta \phi$, the corresponding intensity modulation may be described by a series expansion about the reference point. To first order, the phase perturbation results in an intensity modulation of
\begin{equation}\label{eq:approx}
I(\phi_0 + \Delta \phi) \approx I(\phi_0) + m(\phi_0)\Delta \phi
\end{equation}
where $m$ is the derivative of intensity with respect to the phase $m(\phi) \coloneqq I'(\phi)$. The function $m(\phi)$ can quantify the local sensitivity of the interferometer. 

A more general concept of phase can be obtained by considering that the phase carried by the optical state need not directly stem from a physical phase perturbation; rather, it might be a function of this. In particular, the phase shift $\phi$ applied in one optical path may undergo a transformation $\gamma$ so that the intensity modulation readout provides direct information about $\gamma(\phi)$, not $\phi$. Applying $\gamma^{-1}$ or an approximation of this to the measured value $\gamma(\phi)$ maps the measured data back to the original space, extracting the physical phase $\phi$. When the magnitude of $d\gamma/d\phi$ evaluated at a given bias point is large, the effective sensitivity is increased, as 
\begin{equation}
m(\phi_0) = \frac{dI(\gamma(\phi))}{d\phi}\bigg |_{\phi_0} = \frac{dI}{d\gamma}\frac{d\gamma}{d\phi}\bigg|_{\phi_0}.
\end{equation}
\textit{We call any device that alters the phase in this way, without altering the intensity and direction of the light, an optical phase amplifier.} The term $d\gamma/d\phi$ can then be viewed as the amplifier gain.

In this article, we develop the concept of optical phase amplification, deriving an example device that is entirely comprised of linear optics. Optical feedback is used to imprint a nonlinearly varying phase on the incident light. The emergence of a nonlinear phase from linear optics is due to the geometric series summation of the round-trip phase. Combining with the one-way symmetry of a beam-splitter allows the device to imprint this modified phase while leaving both the intensity and direction unchanged. This approach to a nonlinear phase response is unlike a standard resonator, such as a Fabry-Perot etalon, whose variation in intensity is larger for sharper phase responses. In addition, the response of these devices usually depends on fixed material constants. In the proposed phase amplifier, its response can be continuously tuned by modifying internal phase shifts, implying it can be used at any operating wavelength.

Linear optical phase amplifiers can be realized in many other configurations. Here we demonstrate the simplest configuration that can obtain a continuously tunable gain $d\gamma/d\phi$ that can be operated at any wavelength of interest. Combining these properties with the fact that these devices can replace any phase shifting node leads to an immense number of potential uses. First, the sharply sloped, high-gain regions generate a large output phase response for a small input phase. This could boost phase sensing resolution. Since the device's bias point and perturbation channel $\phi$ reside in a loop, amplified Sagnac phases can be sensed when the apparatus is rotated \cite{RevModPhys.39.475}. Increased phase response could also be used to enhance the light-matter interactions\cite{Chmielak:11, Liu2022, Lin:23, Li2020}. This could produce faster switching times, a smaller half-wave voltage $V_{\pi}$, and/or a smaller physical footprint. Often photonics platforms are designed around materials exhibiting a large optical response to an external perturbation  \cite{Zhu:21, BTO}. Using a sufficiently high-gain phase amplifier, materials with a weaker optical response but other advantages might be considered. 

The flat-sloped regions of the amplifier also carry advantages. These regions offer a smaller phase response for a larger input phase. This could be used to improve stepping precision of a phase controller. Fluctuations in the input phase would also be proportionally reduced, leading to increased stability. Overall, the ability to freely tune between sharp and flat-sloped regions is very powerful, as this can be used to infer large phase shifts that would normally land in a different cycle than the bias point. In phase contrast microscopy \cite{pc2, pc1}, biological samples often fall outside the unambiguous range \cite{Kurata:24}. The tunable gain in the phase amplifier could provide an alternative means to inferring the phase of such samples. Moreover, with multiple wavelengths, linear optical phase amplifiers act as non-reflective, tunable spectral filters, modifying the dispersive properties of light passing through. A deeper analysis of this mode of operation is forthcoming.

Other approaches exist for enhancing the optical response for purposes such as sensing or modulation \cite{res,Sacher,Wang:18,Chatzitheocharis:23,Jin2019,Mercante:16,Xue:22,Gevorgyan}. These approaches often rely on enhancing the intensity directly; in contrast, a phase amplifier is a phase-controlled source of phase, offering much greater generality. Other techniques can considerably more complex, leading to lower output efficiency. For example, nonlinear interferometry \cite{PhysRevLett.128.033601,Ou} considers the use of optical parametric amplifiers to obtain a signal gain higher than that of the accompanying noise. Another approach uses cascaded nonlinear harmonic generation \cite{Li2022}. 

While these approaches are considerably more complex and do not offer tunable gain at any wavelength, they can globally amplify the phase response, which is typical for a change in optical frequency. In the presented linear scheme, this frequency is fixed. Thus, the amplification occurs locally, exhibiting a tradeoff between sensitivity and dynamic range. This is highly advantageous for applications that use the flat-sloped regions, and is a common situation in other linear amplification systems: higher magnification means smaller field of view in optical instruments, and higher gain in an electronic amplifier implies a lower bandwidth. 
\begin{figure}[ht]
\centering\includegraphics[width=.5\textwidth]{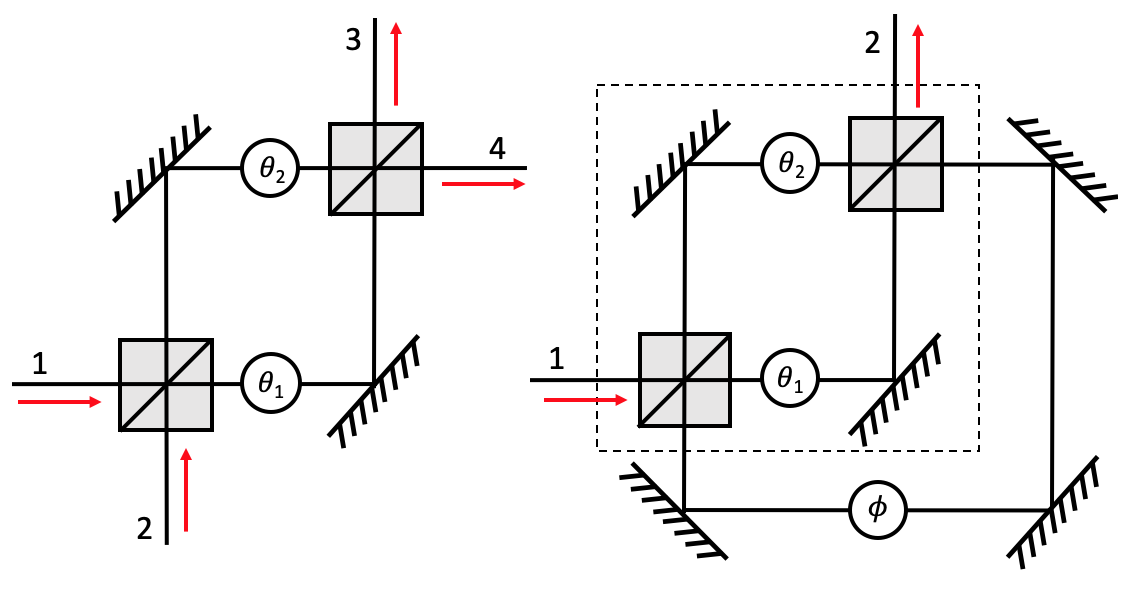}
\caption{(Left) Mach-Zehnder interferometer. (Right) An optical phase amplifier, formed by looping together two ports of a Mach-Zehnder. The device only modifies the phase of an optical state passing through it, and when placed in an external readout, can obtain slopes much larger in magnitude than a standard Mach-Zehnder interferometer. The arrows illustrate how light moves through these feed-forward devices; these could be flipped to correspond to input from the other side.
\label{fig:interferometers}}
\end{figure}

Some phase enhancement techniques use quantum sources since these can have a lower noise power than a classical coherent state \cite{SCHNABEL20171,doi:10.1080/00107510802091298,PhysRevD.23.1693}. In the optimal setting, the uncertainty in the measurement scales quadratically more favorably with the mean value of photon number $N$ \cite{PhysRevA.55.2598}. For example, consider a state of $N$ photons in a balanced superposition of two clusters:
\begin{equation}
|\psi \rangle = \frac{1}{\sqrt{2}} (|N 0\rangle + |0 N\rangle).
\end{equation}
This state is known as a ``N00N state'' \cite{doi:10.1080/00107510802091298}. Each cluster populates the same spatial mode, so when a phase shift $\phi$ is applied within that mode, each photon acquires a phase shift of $\phi$. Collectively, the state sees a phase shift of $N\phi$, obtaining an $N$-fold amplification of the interferogram frequency, obtaining a larger maximum slope and thus higher sensitivity. However, these sources are difficult to efficiently create, especially as $N$ increases, require a very low signal level, and are highly sensitive to losses. 

The device here can realize phase amplification using linear optics and classical light. Although this means the device will not obtain quantum-limit uncertainty scaling, it does allow for probing the phase in a practical manner, utilizing a very high average photon number and single detector readout.

\section{Linear-optical phase amplifier}

Linear optical scattering devices redistribute an optical state among orthogonal field modes. These so-called multiports can be viewed mathematically as a linear transformation acting on the probability amplitudes of the state. Here, we work with states of monochromatic radiation. Unitary scattering devices can be identified with a unitary matrix by identifying the Fock basis creation operator $a^\dagger_j(\mathbf{k})$ of wave-vector $\mathbf{k}$ and spatial mode $j$ with the standard basis vector $\mathbf{e}_j$.

Under this formalism, a 50:50 beam-splitter could be described by the scattering matrix
\begin{equation}\label{eq:bs}
B = \frac{1}{\sqrt{2}}
\begin{pmatrix}
0 & 0 & 1 & 1\\
0 & 0 & 1 & -1\\
1 & 1 & 0 & 0\\
1 & -1 & 0 & 0
\end{pmatrix}.
\end{equation}
Other common conventions could be used. The beam-splitter is sparse in that its ports cannot simultaneously be used for input and output: the diagonal entries are zero. Such a device is called directionally-biased or feed-forward, since light entering a given port cannot reflect out from that same port.

Another device which possesses this feed-forward symmetry, inheriting it from the beam-splitter, is the Mach-Zehnder interferometer. This is pictured in Fig. \ref{fig:interferometers} (left). By looping together two ports we form the feed-forward, $U(2)$ scattering device shown in Fig. \ref{fig:interferometers} (right). These properties imply the following scattering matrix structure
\begin{equation}\label{eq:gen}
U = 
\begin{pmatrix}
     0 & e^{i\gamma_1}\\
     e^{i\gamma_2} & 0
\end{pmatrix}
\end{equation}
where $\gamma_j$ will depend on $\phi, \theta_1$ and $\theta_2$. Because this $U(2)$ device is feed-forward, all incident light must pass through the device. The intensity cannot change but the phase can, and so the device is a phase amplifier. The exact $\gamma_j$ obtained depends on which two ports are used to form the loop. The other cases lead to either the same behavior or a constant output phase.

The computation procedure of $\gamma_1$ and $\gamma_2$ are the same so only a common $\gamma$ will be computed here. We assume both beam-splitters take the form of Eq. (\ref{eq:bs}); using arbitrary and different beam-splitter scattering matrices is similar and is discussed in Appendix A. Defining $\theta \coloneqq \theta_1 - \theta_2$, the scattering matrix of the Mach-Zehnder can be expressed as
\begin{equation}\label{eq:MZI}
U = e^{i\theta/2 + i\theta_2}
\begin{pmatrix}
     \cos(\theta/2) & i \sin(\theta/2)\\
     i\sin(\theta/2) &\cos(\theta/2)\\
\end{pmatrix} \coloneqq
\begin{pmatrix}
     r & t \\
     t & r
\end{pmatrix}.
\end{equation}
Here we use a compressed notation for scattering matrices which possess a certain permutation symmetry. With reference to the port labels in Fig. \ref{fig:interferometers}, the $2\times 2$ form maps excitations in ports (1, 2) to ports (3, 4) and vice versa.

\begin{figure}[ht]
\centering\includegraphics[width=.52\textwidth]{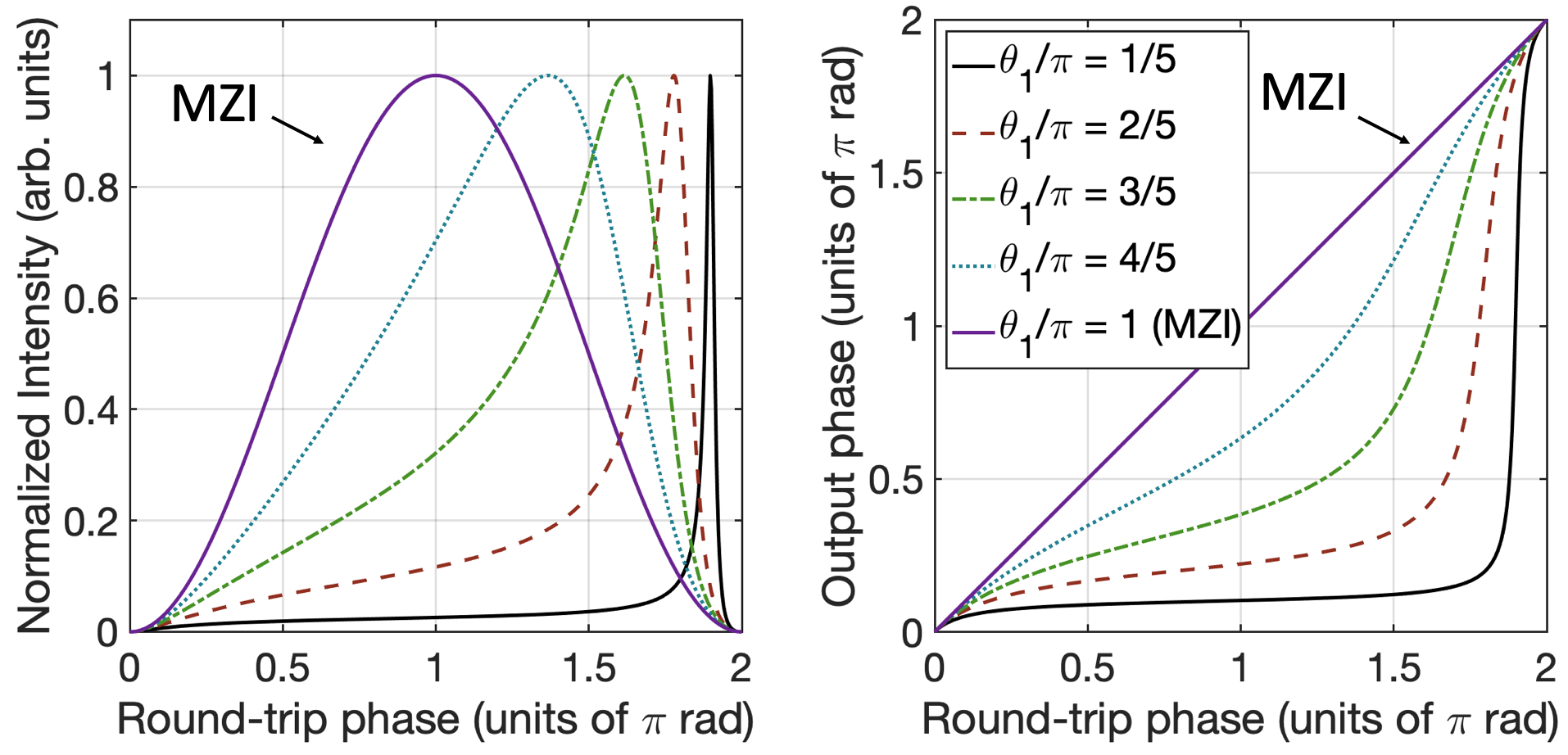}
\caption{Example output of the lossless phase-amplifier with $\theta_2$ set to 0. The intensity (left) is found by overlapping the phase of the device output (right) with a reference beam of constant intensity and phase.\label{fig:theory}}
\end{figure}

When light is incident on the phase amplifier, some light exits the device through the other port and some enters the loop. This feeds back to the first beam-splitter, forming a directionally-biased optical cavity that carries an aggregate round-trip phase of $\phi$. The feed-forward symmetry of the beam-splitter ensures no light transfers from the ring cavity to the port from which the light initially entered the device. The process continues \textit{ad infinitum}, with an output amplitude given by
\begin{align}\label{eq:ffpa}
e^{i\gamma} &= r + t^2e^{i\phi} \sum_{N = 0}^{\infty} (re^{i\phi})^N = r + \frac{t^2e^{i\phi}}{1 - re^{i\phi}}.
\end{align}
A direct formula for $\gamma$ is readily found by expanding Eq. (\ref{eq:ffpa}) in terms of $\theta_1$ and $\theta_2$, separating into real and imaginary parts and finally taking the inverse tangent of their quotient. Each value of $\theta_2$ produces a $\theta$-indexed homotopy $\gamma_{\theta}(\phi)$. For example, when $\theta_2$ = 0,
\begin{equation}\label{eq:h2}
    \gamma_{\theta}(\phi) = \arctan \bigg ( \frac{\sin(\theta) +\cos^2(\theta/2)\sin(\phi) - \sin(\phi + \theta)}{\cos^2(\theta/2)(2 - \cos(\phi))  - \cos(\theta + \phi)}\bigg ).
\end{equation}
If the internal beam-splitter S-matrix convention changes, this and other formulas will vary, but the phases produced will only be changed by an additive constant. Each homotopy shares the property of being a deformation of a line $\gamma = \phi$ into a step function. Example curves for this case are shown in Fig. \ref{fig:theory} beside the intensity modulation they would induce if applied in one arm of a standard Mach-Zehnder interferometer.

We emphasize this is only one among many instances of a phase amplifier. Utilizing feed-forward symmetry in an optical cavity makes it straightforward to derive other phase-amplifying configurations, each with their own particular amplification properties.
\begin{figure}[ht]
\centering\includegraphics[width=.45\textwidth]{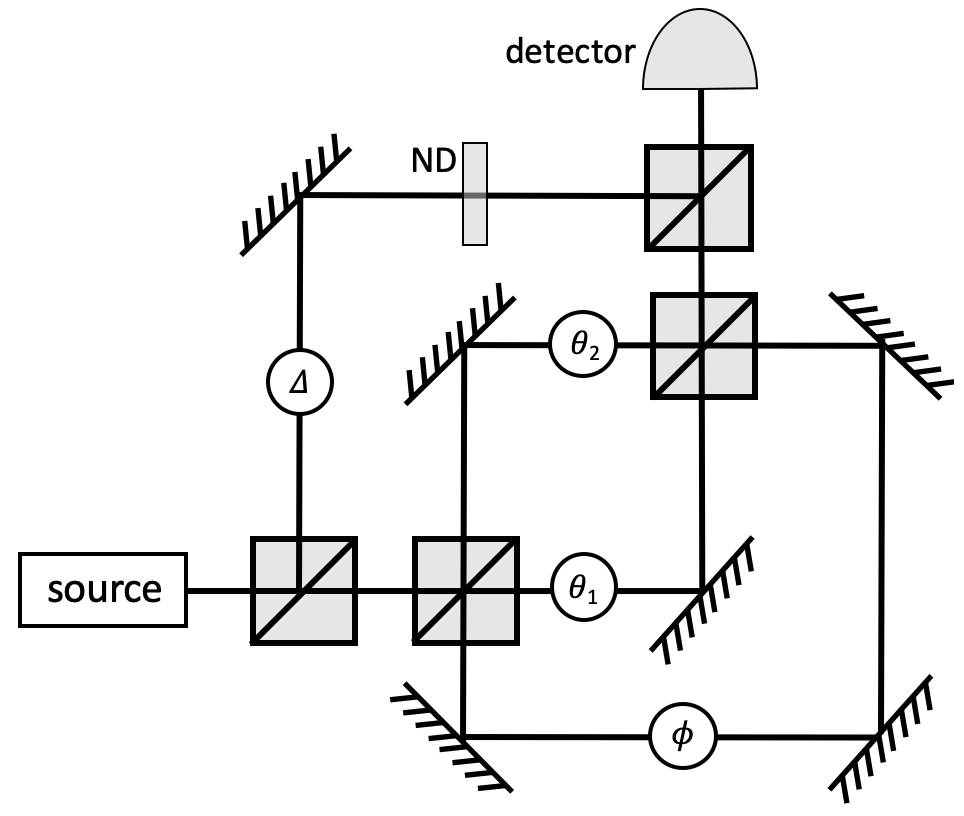}
\caption{Experimental test configuration for the linear optical phase amplifier, used to read out the phase it imparts.\label{fig:setup}}
\end{figure}
\section{Experimental Results}
A phase amplifier has constant intensity with respect to its internal phase parameters. To extract its phase transformation, the device needs to be placed in a readout. Here the phase amplifier was nested recursively in another Mach-Zehnder interferometer. The experimental configuration used is shown in Fig. \ref{fig:setup}. Internal phase shifts were realized with standard optical delay lines with piezoelectric control. A variable neutral-density (ND) filter was placed in the reference arm to vary the intensity of the reference beam. The source was a low power laser with 632.8 nm wavelength. This source was highly coherent, with a spectral linewidth cited at less than 100 kHz. The detector was a standard optical power meter, conducting onboard averaging of data ($N = 15$) sampled at a rate of 1 kHz.

This experimental configuration of Fig. \ref{fig:setup} possesses a property allowing one of the three non-loop phases to be fixed without changing the space of curves observable at the output detector. This is further detailed in Appendix B, and was used to fix the phase of the reference arm in the readout Mach-Zehnder, labeled $\Delta$ in Fig. \ref{fig:setup}. The three piezoelectric crystals were swept through drive voltages in an open loop configuration to span the other parameters.

In an experiment it is natural for some ideal assumptions to break down. In this case, the beam-splitters were not identical nor were they 50:50 and lossless. Mirrors also contribute to losses, accumulating quickly as mirrors are used in succession. With any amount of loss, the output is no longer only a phase transformation, because this would require all of the energy to emerge at the other side of the device. To understand how this affects the phase amplifier, losses were instilled in the model of Eq. (\ref{eq:ffpa}) by allowing the individual beam-splitters to be non-unitary, as well as allowing the loop phase to carry an imaginary component to represent round-trip losses. Numerical studies were conducted with these models to investigate the nature of these deviations from ideal behavior. Values for these loss parameters were derived directly from experimental measurements of the round-trip intensity loss and beam-splitter classical scattering probabilities. Further details are discussed in Appendix A. 

\begin{figure}[ht]
\centering\includegraphics[width=.45\textwidth]{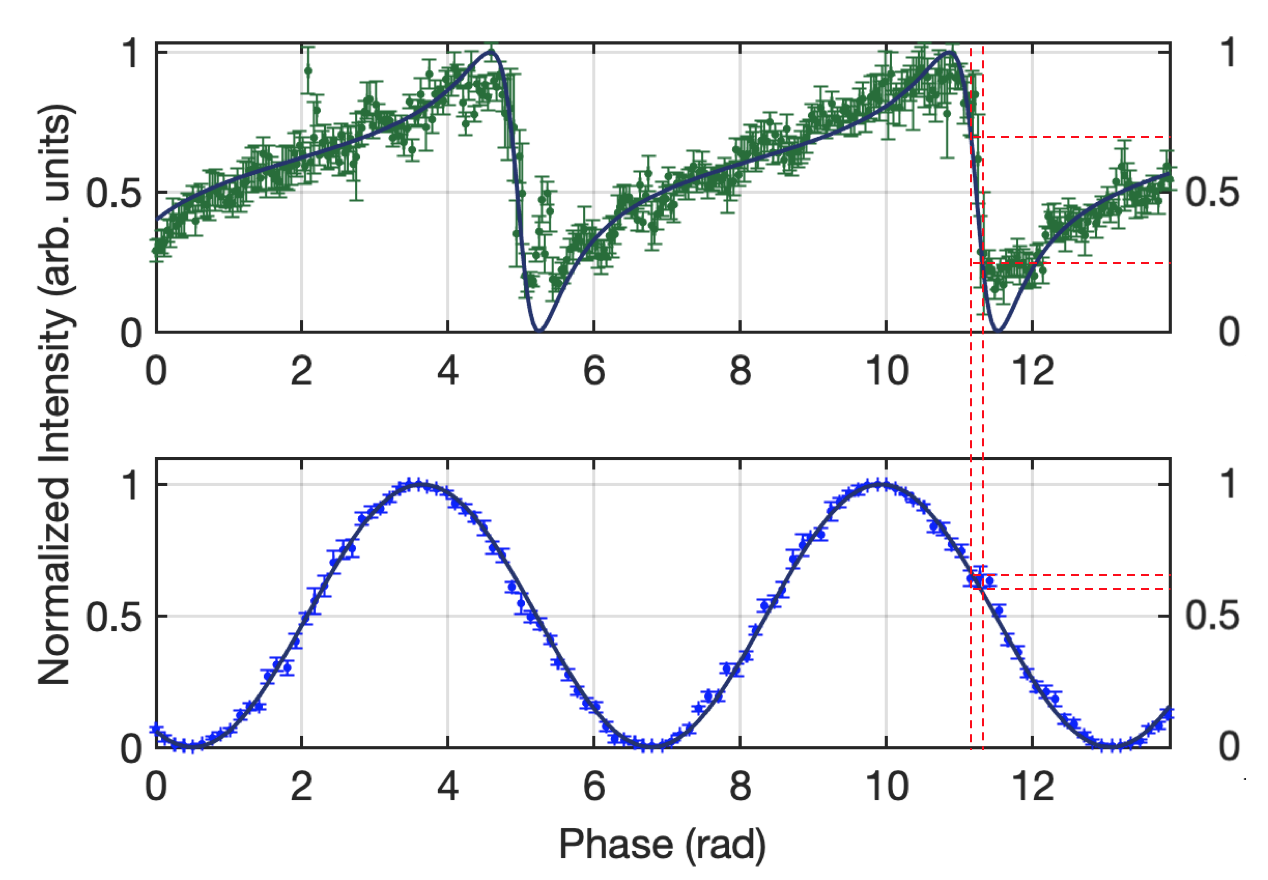}
\caption{Measured curves of a standard (bottom) and phase-amplified (top) Mach-Zehnder interferometer. The phase-amplified version can be biased around a point with a larger slope, which in this case is $m = -10.3 \pm 6.1$. This represents a value more than 20 times steeper than the maximum value of $\pm 1/2$ that can be obtained in the regular Mach-Zehnder interferometer. Dashed lines are shown emanating from the fitted curves to loosely depict the difference in modulation for the same phase perturbation.
\label{fig:data3}}
\end{figure}

Three-phase sweeps were made to the output of the full device in Fig. \ref{fig:setup}. Homotopic evolution of the interference pattern $I(\phi)$ was observed, which went from the traditional Mach-Zehnder linear phase to increasingly nonlinear phase in agreement with Fig. \ref{fig:theory}. The data was fit by finding values for $\theta_1, \theta_2$ and a constant offset to $\phi$ which fit the data best. The amount of optical power in the reference beam affects the slope and visibility at the final output. The optimal value for this amount was found to depend on the amount of loss present in the system. The numerical models suggested that using the ND filter to set the reference beam to lie at roughly 80\% of the power of the maximum phase amplifier output would permit states that possessed good visibility and a large slope. This value was implemented in the experiment and fixed in all model fits.  

In Fig. \ref{fig:data3} data from the internal Mach-Zehnder interferometer is compared to a curve from the phase amplified version. The phase amplified curve possesses a slope of $m = -10.3 \pm 6.1$, representing a phase gain factor of greater than 20 in comparison to the maximum value of $m = \pm 0.5$ that can be obtained with a traditional Mach-Zehnder device. In this case the visibility is slightly above 75\%.

For the device to function properly, the coherence length of the source must be much larger than a round-trip through the loop and one arm of the Mach-Zehnder. In addition to this, the detector must be unable to resolve times between these individual round-trips. When this holds, the probability amplitudes for each path through the system are coherently summed. With our experimental setup having a round-trip distance $\ell < 1$ m, we used a source with coherence length $\ell_c > 10^3$ m and a quasi-DC optical power meter to ensure these conditions were met.

Aligning the phase amplifier device in free space requires simultaneous alignment of a ring cavity and two inputs of a Mach-Zehnder interferometer. Imperfections in this three-faceted alignment are likely the cause for the data to exhibit lower visibility than the model predicts. The transition of the demonstrated device to an integrated waveguided platform will ease or eliminate this issue. The many-beam superposition is also responsible for the comparatively increased fluctuations in the phase amplifier output shown in Fig. \ref{fig:data3}; being only a combination of two beams, the Mach-Zehnder is naturally more stable. The precision of the piezoelectric actuators was measured to be about 1.5 mrad; the corresponding error bars were too small to appear in the figure.

In summary, we have defined and experimentally demonstrated the general principle of linear-optical phase amplification. Devices belonging to this class have the potential for many uses. Besides this, we highlight our perspective that entire interferometers and the symmetries they possess can be used as the basic blocks used to create another device.

\section*{Acknowledgments}
This research was supported by the Air Force Office of Scientific Research MURI Award No. FA9550-22-1-0312, DOD/ARL Grant No. W911NF2020127, and the Beckman Young Investigator Award, from the Arnold and Mabel Beckman Foundation.

\widetext
\section*{Appendix A: non-unitary linear optical phase amplifier}
Here we discuss a model for how the phase amplifier behaves in lossy environments. First, returning to equation Eq. (\ref{eq:ffpa}) for the output amplitude $b$, we have
\begin{align}\label{eq:ffpa2}
b &= r + t^2e^{i\phi} \sum_{N = 0}^{\infty} (re^{i\phi})^N = r + \frac{t^2e^{i\phi}}{1 - re^{i\phi}},
\end{align}
where $r, t$ are the scattering coefficients of the Mach-Zehnder portion and $\phi$ is the loop phase acquired per round-trip. 

If we next assume the scattering coefficients for the beam-splitters comprising the Mach-Zehnder portion are $r_1, t_1$ and $r_2, t_2$, this expression can be rewritten in terms of these coefficients and the arm phases $\theta_1$ and $\theta_2$, which are pictured in Fig \ref{fig:setup}. In this form,
\begin{equation}
    b(\phi, \theta_1, \theta_2) = (t_1 t_2 e^{i\theta_1} + r_1 r_2 e^{i\theta_2}) + \frac{e^{i\phi}(t_1 r_2 e^{i\theta_1} + r_1 t_2 e^{i\theta_2})(r_1 t_2 e^{i\theta_1} + t_1 r_2 e^{i\theta_2})}{1 - e^{i\phi}(r_1 r_2 e^{i\theta_1} + t_1 t_2 e^{i\theta_2})}.
\end{equation}
To model losses, the phases can be complex-valued, and the internal beam-splitter coefficients $r_1, t_1$ and $r_2, t_2$ can be derived from a non-unitary transformation, but we assume for simplicity that this transformation is reciprocal and feed-forward. In particular, we take on the following scattering transformation for the $j$th beam-splitter
\begin{equation}
B_j = 
\begin{pmatrix}
    0 & 0 & i\delta_j & \sqrt{1 - \delta_j^2 - \epsilon_j}\\
    0 & 0 & \sqrt{1 - \delta_j^2 - \epsilon_j} & i \delta_j\\
    i\delta_j & \sqrt{1 - \delta_j^2 - \epsilon_j} & 0 & 0\\
    \sqrt{1 - \delta_j^2 - \epsilon_j} & i \delta_j & 0 & 0\\
\end{pmatrix},
\end{equation}
where $\delta_j$ and $\epsilon_j$ are measured experimentally. Note that any departure from reciprocity would only affect the output of the phase amplifier when the source and detectors were swapped. Since these were kept fixed, it has no effect on the experiment. Cube beam-splitters are known to possess back-reflections, which if not perfectly back-aligned, exit the system. Those losses are encapsulated within $\delta_j$ and $\epsilon_j$. There was also no need to measure the imaginary components of $\theta_1$ and $\theta_2$ because those losses could also be capture by $\delta_j$ and $\epsilon_j$. However, $\text{Im}(\phi)$ was determined from measurements of intensity at the beginning and end of a single round trip. Their ratio $\alpha$ could be converted to $\text{Im}(\phi)$ with the relation $\text{Im}(\phi) = -\log (\sqrt{\alpha} )$. From this one can verify the round-trip amplitude attenuation factor is $\exp(i\cdot i\text{Im}(\phi)) = \exp(\log(\sqrt{\alpha})) = \sqrt{\alpha}$, corresponding to an intensity attenuation factor of $\alpha$ each round-trip. 

The final intensity $I$ collected at the detector of the system in Fig. (\ref{fig:setup}) was modeled as a balanced overlap between the reference beam and phase amplifier output
\begin{equation}\label{eq:intensity}
    I = \frac{1}{\sqrt{2}} | b + \sqrt{A}e^{i\Delta} |^2.
\end{equation}
The amplitude $A$ was set using the neutral density filter, and the variable $\Delta$, being redundant, could be fixed at any value. After fixing values for $\delta_1, \delta_2, \epsilon_1, \epsilon_2$, $\alpha$, and $A$ from lab measurements and setting $\Delta = 0$, we fit our measured data to the equation above using a basic form of maximum likelihood estimation: the remaining degrees of freedom, namely the values of $\theta_1$ and $\theta_2$, as well as a constant immeasurable offset to $\text{Re}(\phi)$, were each selected by finely sweeping $[0, 2\pi]$ and choosing the values which led to the best fit. A similar procedure was used for fitting the data collected from a conventional Mach-Zehnder interferometer, except that only one parameter had to be fit. Both the measured and model curves were normalized prior to fitting to reduce the number of fit parameters and to always provide a fair basis to compare the slope, which scales with maximum intensity. Generally, the fits captured the curves and their expected slope enhancements well, building confidence in the accuracy of the measured parameters. However, the fits failed to capture the visibility, which we attribute to the fact that the modeling does not to account for potential misalignment between different non-planar beams.

The present model for losses was is useful for obtaining insight on the expected tradeoff between the maximum amplifier gain $ d \arg b / d \text{Re}(\phi)$ and its signal level $|b|^2$ vs. the aggregate losses in the system. This tradeoff can be most easily understood when the Mach-Zehnder is treated ideally, so all loss occurs in the loop. Returning to Eq. (\ref{eq:ffpa2}), we see that as the loop losses increase, higher-order terms in the geometric series decay much faster than without any loss present. These terms are only relevant when $|r|^2$ is close to 1. This corresponds to a high-finesse loop cavity that exhibits a large maximum phase-to-phase slope. As a result, these sharply sloped phase regions to exhibit comparatively low robustness to losses. With a fairly high-finesse ring cavity, for instance, $|r|^2 = 0.9$, this model predicts that intensity losses of $2\%$ percent per round-trip lead to a final output attenuation factor of $0.46$ at the sharpest-sloped point. However, this attenuation region is very narrow, so that the flat-sloped regions are barely affected, losing much less than $1\%$ in the majority of each modulation period. How significantly these losses manifest in a real system, especially in comparison to nonlinear and quantum phase amplification techniques, remains an area of active investigation.

\section*{Appendix B: phase amplifier readout parameter reduction}

With the formalism established above, one can derive a useful reduction that allows the removal of one of the four phase parameters from consideration. In particular any one of the three parameters $\Delta, \theta_1, \theta_2$ can be discarded. To that end, first recall that $\theta_1$ and $\theta_2$ possess a well-known exchange symmetry, so that in what follows one can always be substituted for the other. So, without loss of generality, shift $\theta_2$ by an arbitrary amount $\alpha$. This has the following effect on the phase amplifier output amplitude $b$: 
\begin{align}
    b(\phi, \theta_1, \theta_2 + \alpha) &= (t_1 t_2 e^{i\theta_1} + r_1 r_2 e^{i\theta_2 + i\alpha}) + \frac{e^{i\phi}(t_1 r_2 e^{i\theta_1} + r_1 t_2 e^{i\theta_2 + \alpha})(r_1 t_2 e^{i\theta_1} + t_1 r_2 e^{i\theta_2 + i\alpha})}{1 - e^{i\phi}(r_1 r_2 e^{i\theta_1} + t_1 t_2 e^{i\theta_2 + i\alpha})} \\ &=
    e^{i\alpha} (t_1 t_2 e^{i\theta_1 - i \alpha} + r_1 r_2 e^{i\theta_2}) + \frac{e^{i\phi + 2i\alpha}(t_1 r_2 e^{i\theta_1-i\alpha} + r_1 t_2 e^{i\theta_2})(r_1 t_2 e^{i\theta_1 - i\alpha} + t_1 r_2 e^{i\theta_2})}{1 - e^{i\phi + i\alpha}(r_1 r_2 e^{i\theta_1 -i \alpha} + t_1 t_2 e^{i\theta_2})} \\ &=
    e^{i\alpha} \bigg ( (t_1 t_2 e^{i\theta_1 - i \alpha} + r_1 r_2 e^{i\theta_2}) + \frac{e^{i\phi + i\alpha}(t_1 r_2 e^{i\theta_1-i\alpha} + r_1 t_2 e^{i\theta_2})(r_1 t_2 e^{i\theta_1 - i\alpha} + t_1 r_2 e^{i\theta_2})}{1 - e^{i\phi + i\alpha}(r_1 r_2 e^{i\theta_1 -i \alpha} + t_1 t_2 e^{i\theta_2})} \bigg) \\ &= b(\phi + \alpha, \theta_1 - \alpha, \theta_2)e^{i\alpha}.
\end{align}
Now we use this as a lemma in conjunction with applying the same shift to the final output intensity in Eq. (\ref{eq:intensity}).
\begin{align}
    I(\phi, \theta_1, \theta_2 + \alpha, \Delta) &= |c_1 \sqrt{I_r}e^{i\Delta}  + c_2 e^{i\alpha}b(\phi + \alpha, \theta_1 - \alpha, \theta_2)|^2 \\ &= 
    |e^{i\alpha} (c_1 \sqrt{I_r}e^{i\Delta - i\alpha}  + c_2 b(\phi + \alpha, \theta_1 - \alpha, \theta_2)|^2 \\ &= |c_1 \sqrt{I_r}e^{i\Delta - i\alpha}  + c_2 b(\phi + \alpha, \theta_1 - \alpha, \theta_2)|^2 \\ &= I(\phi - \alpha, \theta_1 + \alpha, \theta_2, \Delta - \alpha)
\end{align}
An immediate corollary of this is that $I(\phi, \theta_1, \theta_2, \Delta + \alpha) = I(\phi + \alpha, \theta_1 - \alpha, \theta_2 - \alpha, \Delta)$. 

The implication is that a shift in $\Delta$ or $\theta_2$ is equivalent to a shift in the other by the opposite amount as well as a shift in the homotopy phases $\phi$ and $\theta_1$. However one can never know $\phi$ and $\theta_1$ globally anyways; so this shift is undetectable. For a given $\Delta, \theta_2$ one obtains a new continuous family of curves $I_{\theta_1}(\phi | \Delta, \theta_2)$. Hence, to view all possible families of curves, one only must sweep one of either $\theta_2$ and $\Delta$, since an arbitrary shift in either can be always recast as a change in the other parameters.

\bibliography{refs}
\end{document}